\magnification 1200
\centerline {\bf On Connections between the Quantum and Hydrodynamical Pictures of 
Matter}
\vskip 0.5cm
\centerline {\bf  by Geoffrey Sewell}
\vskip 0.5cm
\centerline {\bf Department of Physics, Queen Mary, University of London,}
\vskip 0.2cm
\centerline {\bf  Mile End Road, London E1 4NS, UK. Email: g.l.sewell@qmul.ac.uk}
\vskip 1cm
\centerline {\bf Abstract} 
\vskip 0.3cm
We  present a general, model-independent,  quantum statistical treatment of the 
connection between the quantum and hydrodynamic pictures of reservoir driven 
macroscopic systems. This treatment is centred on the large scale properties of locally 
conserved hydrodynamical observables and is designed to provide a bridge between 
quantum microdynamics and classical macroscopic continuum mechanics, rather than a 
derivation of the latter from the former. The key assumptions on which the treatment is 
based are hypotheses of chaoticity and local equilibrium for the hydrodynamical 
fluctuations around nonequilibrium steady states, together with an extension of 
Onsager\rq s regression hypothesis to these states. On this basis, we establish canonical 
generalisations of both the Onsager reciprocity relations and the Onsager-Machlup 
fluctuation theory to nonequilibrium steady states, and we show that the spatial 
correlations of the hydrodynamical fluctuations are generically of long range in these 
states.
\vskip 1cm
\centerline {\bf 1. Introduction}
\vskip 0.3cm
It is an empirical fact that the laws of classical macroscopic continuum mechanics are 
extremely general and independent, in form, of  microscopic constitution. This suggests 
that it should be possible to base a statistical mechanical treatment of these laws on very 
general arguments that are centred on macroscopic observables of the hydrodynamical 
type. In fact, such a treatment was initiated by Onsager [1], in his ground-breaking work 
of nonequilibrium thermodynamics, which served to relate the macroscopic dynamics of 
systems close to equilibrium to certain extremely general properties of their underlying 
microscopic dynamics. This kind of approach to the connection between nonequilibrium 
thermodynamics and microphysics has been subsequently pursued by the present author 
in works [2-4] designed to form a bridge between quantum microdynamics and 
macroscopic quantum continuum mechanics rather than a derivation of the latter from the 
former\footnote*{It is thus radically different from the derivations of Euler, though not 
Navier-Stokes, hydrodynamics of models of a plama [5] and of a fermionic system with 
short range intercations [6] from their underlying quantum dynamics}: its scope, 
moreover, is not restricted to states close to global equilibrium. 
\vskip 0.2cm
The present article is devoted to an expositary account of our quantum macrostatistical 
picture of  the nonequilibrium thermodynamics of reservoir driven macroscopic systems 
[3, 4]. This is based on general assumptions of (a) local equilibrium at a certain 
mesoscopic level,  (b) chaoticity of the currents associated with the locally conserved 
quantum fields and (c) an extension  of Onsager\rq s regression hypothesis to fluctuations 
about nonequilibrium states. On the basis of these hypotheses, which have been 
substantiated for certain tractable models [7], we have obtained both a nonlinear 
generalisation of Onsager\rq s theory and the result that the spatial correlations of 
hydrodynamical observables of reservoir driven open systems are generically of long 
range in nonequilibrium steady states. This latter result, which had previously been 
obtained for some special classical stochastic models [8-10], marks a crucial difference 
between equilibrium and nonequilibrium states, since the spatial correlations carried by 
the former are of short range, except at critical points. 
\vskip 0.2cm
In Section 2, we present our model of macroscopic quantum systems in general terms, 
from both the microscopic and the hydrodynamical standpoints. In Section 3, we 
formulate the fluctuation process executed by the hydrodynamical variables, subject to 
the above assumptions (a)-(c), and from these we derive canonical generalisations of both 
the Onsager reciprocity relations and the Onsager-Machlup [11] fluctuation dynamics to 
nonequilibrium steady states. In Section 4 we show that, under the above assumptions, 
the spatial correlations of hydrodynamical observables  in nonequilibrium steady states 
are generically of long range. We conclude in Sec. 5 with further brief comments on the 
theory presented here. 
\vskip 0.5cm
\centerline {\bf 2. The Model}
\vskip 0.3cm
We take the model to be a quantum system, ${\Sigma}$, of $N$  particles located in a  
bounded open connected region ${\Omega}_{N}$ of a $d$-dimensional space, $X$, and 
coupled at its boundary, ${\partial}{\Omega}_{N}$, to an array of reservoirs 
${\cal R}={\lbrace}{\cal R}_{\alpha}{\rbrace}$. ${\Sigma}$ is thus an open system, 
comprising part of the composite ${\Sigma}^{c}:=({\Sigma}+{\cal R})$, which we 
assume to be conservative. The particle number  $N$ is a variable parameter of the 
model. We assume that ${\Omega}_{N}$ is the dilation by a factor $L_{N}$ of a fixed 
region ${\Omega}$ of unit volume and that both ${\Omega}$ and the mean particle 
number density, ${\nu}$, of ${\Sigma}$ are $N$-independent. Thus 
${\Omega}_{N}={\lbrace}L_{N}x{\vert}x{\in}{\Omega}{\rbrace}$ and 
${\nu}L_{N}^{d}=N$. Throughout this article we employ units in which ${\hbar}$ and 
$k_{Boltzmann}$ are equal to unity.
\vskip 0.3cm
\centerline {\bf 2.1. The Quantum Statistical Picture.}
\vskip 0.3cm
We employ the standard operator algebraic description [12-14, 2] of  conservative system 
such as ${\Sigma}^{c}$ or ${\Sigma}$, in the situation where it is isolated from ${\cal 
R}$. Thus, we recall that, in this description, the observables of such a system are the 
self-adjoint elements of a $^{\star}$-algebra, ${\cal A}$, its states, ${\omega}$, are 
linear, positive, normalised, expectation  functionals 
$A{\rightarrow}{\langle}{\omega}:A{\rangle}$ on ${\cal A}$, and its dynamics, in the 
Heisenberg picture, corresponds to a one-parameter group of automorphisms 
$A{\rightarrow}A_{t}$ of ${\cal A}$.We assume that all its interactions are invariant 
under space translations, rotations and time reversals. The equilibrium states of the 
system at inverse temperature ${\beta}$ are characterised by the Kubo-Martin-Schwinger 
(KMS) condition [15, 2]:-
$${\langle}{\omega}:A_{t}B{\rangle}={\langle}{\omega};BA_{t+i{\beta}}{\rangle}, \ 
{\forall} \ A,B{\in}{\cal A}.$$ 
The set of states satisfying this condition is convex, and its extremal elements correspond 
to pure thermodynamic phases [16, 2].
\vskip 0.2cm
We assume that ${\Sigma}$ has a set of extensive, conserved observables 
${\hat Q}=({\hat Q}_{1},. \ .,{\hat Q}_{n})$, which intercommute up to surface 
corrections and are {\it thermodynamically complete} in the sense that the equilibrium 
states corresponding to pure phases are labelled by the expectation values of the global 
density of ${\hat Q}$ in the limit $N{\rightarrow}{\infty}$ [2]. We denote by $s(q)$ the 
equilibrium entropy density for which this expectation value is $q=(q_{1},. \ .,q_{n})$. 
The function $s$ is concave and  may  be formulated by standard statistical mechanical 
procedures [12, 2]. The thermodynamical control variable conjugate to $q$ is then 
${\theta}=({\theta}_{1},. \ .,{\theta}_{n})$, with 
${\theta}_{r}={\partial}s(q)/{\partial}q_{r}$. Thus, assuming that ${\hat Q}_{1}$ is the 
energy observable, ${\theta}_{1}$ is the inverse temperature. We restrict our 
considerations to situations where the system is confined to a single phase region for 
which $q$ lies in a connected domain, ${\Delta}$, of ${\bf R}^{n}$, in which the 
function $s$ is smooth and the correspondence between $q$ and ${\theta}$ is one-to-one. 
It follows that the Hessian matrix $s^{{\prime}{\prime}}(q) \ 
:=[{\partial}^{2}s/{\partial}q_{k}{\partial}q_{l}]$  is invertible and we define
$$J(q):=-s^{{\prime}{\prime}}(q)^{-1}.\eqno(2.1)$$ 
Further, by the one-to-one correspondence between $q$ and ${\theta}$, the equilibrium 
states is this phase may equivalently be labelled by the latter or the former. We assume 
the symmetries of time inversions, space translations and space rotations are unbroken in 
this pure equilibrium phase. 
\vskip 0.3cm
{\it The Reservoirs.} We assume that each reservoir ${\cal R}_{\alpha}$ has a 
thermodynamically complete set, ${\hat Q}$, of extensive conserved observables 
${\hat Q}_{\alpha}=({\hat Q}_{{\alpha},1},. \ .,{\hat Q}_{{\alpha},n})$, which are the 
natural counterparts of  the ${\hat Q}_{r}$\rq s and which satisfy the condition that the 
${\Sigma}-{\cal R}_{\alpha}$ interactions conserve each 
$({\hat Q}_{r}+{\hat Q}_{{\alpha},r})$. The thermodynamical control variable, 
${\theta}_{\alpha}$, of  ${\cal R}_{\alpha}$ is then the canonical counterpart of the 
variable ${\theta}$ of ${\Sigma}$. We denote by 
${\omega}_{\alpha}({\theta}_{\alpha})$ the equilibrium state of ${\cal R}_{\alpha}$ 
corresponding to ${\theta}_{\alpha}$. 
\vskip 0.3cm
{\it The Steady State of ${\Sigma}^{c}$.} We assume that the system ${\Sigma}$ and 
the reservoirs ${\lbrace}{\cal R}_{\alpha}{\rbrace}$ are independently prepared, in the 
remote past, with ${\Sigma}$ in an arbitrary state ${\phi}$ and each ${\cal R}_{\alpha}$ 
in its equilibrium state ${\omega}_{\alpha}({\theta}_{\alpha})$, and that the reservoirs 
are then coupled to spatially disjoint regions of  ${\partial}{\Omega}_{N}$, whose union 
comprises that surface. Then, under rather general conditions [17, 18], ${\Sigma}^{c}$ 
evolves to a terminal state ${\omega}^{c}$, which evidently depends on the variables 
${\lbrace}{\theta}_{\alpha}{\rbrace}$ In the special case where the 
${\theta}_{\alpha}$\rq s for the different reservoirs are all equal, ${\omega}^{c}$ is a 
canonical equilibrium state. Otherwise, it is the nonequilibrium steady state of 
${\Sigma}^{c}$ for the specified conditions. 
\vskip 0.3cm
{\it  The Local Density of ${\hat Q}$.} We assume that the extensive conserved 
observables ${\hat Q}$ of ${\Sigma}$ have  locally conserved, position dependent 
densities 
${\hat q}(x)=\bigl({\hat q}_{1}(x),. \ .,{\hat q}_{k}(x)\bigr)$ with associated currents 
${\hat j}(x)=\bigl({\hat j}_{1}(x),. \ .,{\hat j}_{n}(x)\bigr)$. More precisely,  we assume 
that the local conservation law for ${\hat q}$ prevails for the evolution of this field not 
only for the dynamics of ${\Sigma}$, when isolated, but also for that of the composite 
system ${\Sigma}^{c}=({\Sigma}+{\cal R})$.  Thus, denoting by 
${\hat q}_{t}(x){\equiv}
{\hat q}(x,t)$ and ${\hat j}_{t}(x){\equiv}{\hat j}(x,t)$ the evolutes of the fields 
${\hat q}$ and ${\hat j}$, respectively, for the system ${\Sigma}^{c}$, these time-
dependent fields and currents satisfy the local conservation law 
$${{\partial}{\hat q}_{t}\over {\partial}t}+{\nabla}.{\hat j}_{t}=0.$$ 
For simplicity, we assume that both the extensive observable ${\hat Q}$ and its position 
dependent density ${\hat q}$ are invariant under time reversals, i.e. velocity reversals.
\vskip 0.2cm
In accordance with the general requirements of quantum field theory [19], we assume that 
the fields ${\hat q}_{t}$ and ${\hat j}_{t}$  are operator valued 
distributions\footnote*{This assumption may easily be seen to be valid in standard 
physical cases. For example, the field representing the position dependent particle 
number density is ${\sum}_{r=1}^{N}{\delta}(x-x_{r})$, where $x_{r}$ is the position 
of the $r$\rq th particle.}.
\vskip 0.3cm
\centerline {\bf 2.2. The Hydrodynamical Description.}
\vskip 0.3cm
We assume that the hydrodynamical picture is given by a continuum mechanical law 
governing the evolution of an $n$-component, locally conserved classical field 
$q_{t}(x)=\bigl(q_{1,t}(x),. \ .,q_{n,t}(x)\bigr)$, on a macroscopic space-time scale that 
we shall presently specify. As we shall see in Section 2.3, this field represents a rescaled  
expectation value of the quantum field ${\hat q}_{t}(x)$, but here we shall be concerned 
with just its phenomenological properties.
\vskip 0.2cm
We assume that the current $j_{t}=(j_{1,t},. \ .,j_{n,t})$ associated with $q_{t}$ 
satisfies a constitutive equation of the form
$$j_{t}(x)={\cal J}(q_{t}:x)\eqno(2.2)$$
where ${\cal J}$ is a functional of the field $q_{t}$ and the position $x$. Consequently, 
by local conservation, $q_{t}$ evolves according to an autonomous law,
$${{\partial}\over {\partial}t}q_{t}(x)={\cal F}(q_{t};x){\equiv}
-{\nabla}.{\cal J}(q_{t};x),\eqno(2.3)$$
subject to boundary conditions that are fixed by the reservoirs\footnote{**}{As discussed 
in [4], these conditions serve to equate the spatial boundary value of the local control 
variable ${\theta}_{t}(x):=s^{\prime}(q_{t}(x))$ with that of the reservoir ${\cal 
R}_{\alpha}$ that is in contact with ${\Sigma}$ at $x$.}. We assume that the resultant 
$q_{t}(x)$ is confined to the single phase region ${\Delta}$, introduced in Sec. 2.1. For 
simplicity, we base our explicit treatment here on the case of nonlinear diffusions where 
$${\cal J}(q;x)=-K\bigl(q_{t}(x)\bigr){\nabla}q_{t}(x); \ {\cal F}(q_{t};x)=
{\nabla}.\bigl(K\bigl(q_{t}(x)\bigr){\nabla}q_{t}(x)\bigr),\eqno(2.4)$$
and where $K(q)$ is an $n$-by-$n$ matrix $[K_{kl}(q)]$ and acts by matrix 
multiplication on ${\nabla}q$ . This evolution is therefore invariant under scale 
transformations $x{\rightarrow}{\lambda}x, \ t{\rightarrow}{\lambda}^{2}t$. 
\vskip 0.2cm
We choose the unit of length for the hydrodynamic scale to be $L_{N}$. Thus, in view 
of our stipulation that ${\Omega}_{N}=L_{N}{\Omega}$, the region occupied by 
${\Sigma}$ in the hydrodynamical picture is ${\Omega}$. Furthermore, the scale 
invariance noted after Eq. (2.4) signifies that a length scale $L_{N}$ corresponds to a 
time scale $L_{N}^{2}$. We assume that  that there is just one hydrodynamically steady 
state of the system, as represented by the field $q(x)$ and the current $j(x)$ for which 
${\cal F}(q;x)=0$ and $j(x)={\cal J}(q;x)$. To lighten the notation, we define
$$K_{q}:=K{\circ}q; \ J_{q}:=J{\circ}q; \ {\tilde K}_{q}:=K_{q}J_{q}.\eqno(2.5)$$
\vskip 0.3cm 
{\it Hydrodynamical Perturbations of the Steady State.} By Eq. (2.3), the linearised 
equations of motion for \lq small\rq\ perturbations ${\delta}q_{t}$ and ${\delta}j_{t}$ of 
the steady state field and current $q(x)$ and $j(x)$, respectively,  are 
$${\delta}j_{t}(x)={\cal K}{\delta}q_{t}(x):=
{{\partial}\over {\partial}{\lambda}}{\cal J}(q+{\delta}q_{t};x)_{{\vert}{\lambda}=0}; 
\  {{\partial}\over {\partial}t}{\delta}q_{t}(x)={\cal L}{\delta}q_{t}(x)
:= {{\partial}\over {\partial}{\lambda}}
{\cal F}(q+{\delta}q_{t};x)_{{\vert}{\lambda}=0}.\eqno(2.6)$$
\vskip 0.2cm
We note that, since $q_{t}$ and $j_{t}$ are rescaled expectation values of the 
distribution valued quantum fields ${\hat q}_{t}$ and ${\hat j}_{t}$, it follows that they 
too are distributions. Specifically, denoting by ${\cal D}({\Omega})$ and 
${\cal D}_{V}({\Omega})$ the L. Schwartz spaces [19] of real, infinitely differentiable 
scalar and vector valued functions, respectively, on ${\Omega}$ with support in the 
interior of that region, we take ${\delta}q_{t}$ and ${\delta}j_{t}$ to be elements of  the 
duals, 
${\cal D}^{{\prime}n}({\Omega})$ and ${\cal D}_{V}^{{\prime}n}({\Omega})$, of 
the $n$-fold tensorial powers, ${\cal D}^{n}({\Omega})$ and ${\cal 
D}^{n}_{V}({\Omega})$, of ${\cal D}({\Omega})$ and ${\cal D}_{V}({\Omega})$, 
respectively. We denote elements of ${\cal D}^{n}({\Omega})$ and ${\cal 
D}^{n}_{V}({\Omega})$ by $f=(f_{1},. \ .,f_{n})$ and $g=(g_{1},. \ .,g_{n})$, 
respectively, and we equip these latter spaces with inner products $(.,.)$ and $(.,.)_{V}$, 
respectively, defined by the formulae
$$(f,f^{\prime})={\sum}_{r=1}^{n}\int_{\Omega}dxf_{r}(x)f_{r}^{\prime}(x) ; \  
 (g,g^{\prime})_{V}={\sum}_{r=1}^{n}\int_{\Omega}dxg_{r}(x).g_{r}^{\prime}(x).
\eqno(2.7)$$
\vskip 0.2cm
According to the above definitions, ${\cal L}$ is a linear transformation of 
${\cal D}^{{\prime}n}({\Omega})$ and ${\cal K}$  is a linear mapping of 
${\cal D}^{{\prime}n}({\Omega})$ into ${\cal D}_{V}^{{\prime}n}({\Omega})$. By 
Eqs. (2.3) and (2.6), these linear operators  are related by the equation
$${\cal L}=-{\nabla}.{\cal K}.\eqno(2.8)$$
We assume that ${\cal L}$ is the generator of a one-parameter semigroup 
${\lbrace}T_{t}{\vert}t{\in}{\bf R}_{+}{\rbrace}=T({\bf R}_{+})$ of linear 
transformations of  ${\cal D}^{{\prime}n}({\Omega})$, and we denote the duals of 
${\cal L}$ and $T_{t}$ by ${\cal L}^{\star}$ and $T_{t}^{\star}$, respectively. Thus, 
by Eq. (2.6),
$${\delta}q_{t}=T_{t-s}{\delta}q_{s}; {\delta}j_{t}=-{\cal K}T_{t-s}{\delta}q_{s} \ 
{\forall} \ t{\geq}s.\eqno(2.9)$$
\vskip 0.3cm
\centerline {\bf 2.3. Connection between the Quantum and Hydrodynamical Pictures.}
\vskip 0.3cm
In order to bring our formulation of the quantum field ${\hat q}_{t}(x)$  and current 
${\hat j}_{t}(x)$ into line with their hydrodynamical counterparts, we rescale the 
position $x$ and the time $t$ by the factors $L_{N}$ and $L_{N}^{2}$, respectively. 
The resultant quantum field ${\tilde q}_{t}(x)$ and current ${\tilde j}_{t}(x)$ are 
therefore given by the formulae 
$${\tilde q}_{t}(x)={\hat q}(L_{N}x,L_{N}^{2}t); \ 
{\tilde j}_{t}(x)=L_{N}{\hat j}(L_{N}x,L_{N}^{2}t).\eqno(2.10)$$
Thus, ${\tilde q}_{t}$ and ${\tilde j}_{t}$ satisfy the local conservation law 
$${{\partial}{\tilde q}_{t}\over {\partial}t}+{\nabla}.{\tilde j}_{t}=0.$$
\vskip 0.2cm 
We assume that, in general, the phenomenological fields $q_{t}$ and $j_{t}$ of Sec. 2.2 
are just the expectation values of ${\tilde q}_{t}$ and ${\tilde j}_{t}$, respectively, for 
the prevailing state of ${\Sigma}^{c}$ in the limit where $N$ tends to infinity. In 
particular, the stationary field $q(x)$  and current $j(x)$ are the limiting values,  as $N$ 
tends to infinity, of ${\langle}{\omega}^{c};{\tilde q}(x){\rangle}$  and 
${\langle}{\omega}^{c};{\tilde j}(x){\rangle}$, respectively.
\vskip 0.3cm
{\it The Hydrodynamical Fluctuation Fields.} We represent the fluctuations of  the 
hydrodynamic field ${\tilde q}_{t}$ and current ${\tilde j}_{t}$ about their means for 
the steady state ${\omega}^{c}$ by the fluctuation field ${\tilde {\xi}}_{t}=
({\tilde {\xi}}_{1,t},. \ .,{\tilde {\xi}}_{n,t})$ and its associated current 
${\tilde {\eta}}_{t}=({\tilde {\eta}}_{1,t},. \ .,{\tilde {\eta}}_{n,t})$, as defined by the 
formulae
$${\tilde {\xi}}_{t}(x)=N^{1/2}\bigl({\tilde q}_{t}(x)-
{\langle}{\omega}^{c};{\tilde q}_{t}(x){\rangle}\bigr); \ 
{\tilde {\eta}}_{t}(x)=N^{1/2}\bigl({\tilde j}_{t}(x)-
{\langle}{\omega}^{c};{\tilde j}_{t}(x){\rangle}\bigr),\eqno(2.11)$$
the factor $N^{1/2}$ being canonical for fluctuations [20]. It follows immediately from 
these definitions that the local conservation law for ${\hat q}_{t}$ implies the 
corresponding one for ${\tilde {\xi}}_{t}$, which we express in the following integral 
form.
$${\tilde {\xi}}_{t}-{\tilde {\xi}}_{s}+{\tilde {\zeta}}_{t,s}=0  \ {\forall} \ t,s{\in}
{\bf R},\eqno(2.12)$$
where
$${\tilde {\zeta}}_{t,s}=\int_{s}^{t}du{\nabla}.{\tilde {\eta}}_{u}.\eqno(2.13)$$ 
\vskip 0.5cm
\centerline {\bf 3.The Fluctuation Process.}
\vskip 0.3cm
We denote by ${\tilde {\xi}}_{t}(f)$ and ${\tilde {\zeta}}_{t,s}(g)$ the time-dependent 
smeared fields obtained by integrating ${\tilde {\xi}}_{t}$ and ${\tilde {\zeta}}_{t,s}$ 
against test functions $f \ ({\in}{\cal D}^{n}({\Omega}))$ and $g \ ({\in}{\cal 
D}_{V}^{n}({\Omega}))$, respectively. The statistical properties of these smeared 
fields in the steady state ${\omega}^{c}$ are then represented by the correlation 
functions given by the expectation values, for this state, of the monomials in the 
${\tilde {\xi}}_{t}(f)$\rq s and the ${\tilde {\zeta}}_{t,s}(g)$\rq s. We assume that these 
functions are continuous in all their arguments (the test functions $f,g$ and the time 
variables $s,t$) and converge, as $N{\rightarrow}{\infty}$, to corresponding finite 
valued ones for a stochastic process executed by  fields 
$\big({\xi}_{t}(f),{\zeta}_{t,s}(g)\bigr)$, respectively: conditions for the validity of 
these assumptions are specified in [4]. Then, in view of the completeness of the Schwartz 
${\cal D}$-spaces, the correlation functions for the resultant process are continuous in 
the test functions $f$ and $g$ and measurable in the time variables. Further, under 
general space-time asymptotic abelian conditions on the fields $\bigl({\tilde 
{\xi}}_{t}(f),{\tilde {\zeta}}_{t,s}(g)\bigr)$, the process is classical [4]. Thus, assuming 
these conditions, the quantum process $({\tilde {\xi}},{\tilde {\zeta}})$ converges {\it in 
law}, as $N$ tends to infinity, to the classical process $({\xi},{\zeta})$. 
It follows immediately that ${\xi}$ and ${\zeta}$ satisfy the natural counterpart of the 
local conservation law (2.12), which when integrated against a test function $f \ 
({\in}{\cal D}^{n}({\Omega}))$ takes the form
$${\xi}_{t}(f)-{\xi}_{s}(f)={\zeta}_{t,s}({\nabla}f).\eqno(3.1)$$ 
\vskip 0.3cm
\centerline {\bf  3.1. The Regression Hypothesis.} 
\vskip 0.3cm
We assume a natural generalisation of Onsager\rq s [1] regression hypothesis to the effect 
that the fluctuation field ${\xi}_{t}$ regresses according to the same dynamical law as 
the weak externally induced perturbations ${\delta}q_{t}$ of the stationary 
hydrodynamical field $q$. Thus, since the latter law is given by Eq. (2.9) , we take our 
regression hypothesis to be that 
$$E({\xi}_{t}{\vert}{\xi}_{s})=T_{t-s}{\xi}_{s} \ {\forall} \ t{\geq}s,\eqno(3.2)$$
where $E(.{\vert}{\xi}_{s})$ denotes the conditional expectation given the field 
${\xi}_{s}$ at time $s$. Hence, denoting the static two-point function by
$$W(f,f^{\prime}):=E\bigl({\xi}(f){\xi}(f^{\prime})\bigr),\eqno(3.3)$$
it follows from Eq. (3.2) and the stationarity of the ${\xi}$-process that 
$$E\bigl({\xi}_{t}(f){\xi}_{s}(f^{\prime})\bigr)=W(T_{t-s}^{\star}f,f^{\prime}) \ 
{\forall} \ f,f^{\prime}{\in}{\cal D}^{n}({\Omega}), \ t,s({\leq}t){\in}{\bf R}.
\eqno(3.4)$$
\vskip 0.2cm
We further develop the relation between fluctuations and externally induced 
hydrodynamical perturbations by noting that, by Eq. (2.6), 
$\int_{s}^{t}du{\delta}j_{u}=\int_{s}^{t}du{\cal K}{\delta}q_{u}$.
Correspondingly, we designate the {\it secular} part of the time-integrated current 
fluctuation  ${\zeta}_{t,s}$ to be $\int_{s}^{t}du{\cal K}{\xi}_{u}$, and we define the 
remainder of ${\zeta}_{t,s}$ to be its {\it stochastic} part, namely
$${\zeta}_{t,s}^{stoc}={\zeta}_{t,s}-\int_{s}^{t}du{\cal K}{\xi}_{u}.\eqno(3.5)$$ 
It follows then from Eqs. (2.8) and (3.5)  that the local conservation law (3.1) is 
equivalent to the following equation.
 $${\xi}_{t}(f)-{\xi}_{s}(f)-\int_{s}^{t}du{\xi}_{u}({\cal L}^{\star}f)=
{\zeta}_{t,s}^{stoc}({\nabla}f).\eqno(3.6)$$
Since ${\cal L}$ is the generator of $T({\bf R}_{+})$, it follows from this equation, after 
some manipulation [4], that the two-point function of ${\zeta}^{stoc}$ is related to that 
of ${\xi}$ according to the formula
$$E\bigl({\zeta}_{t,s}({\nabla}f){\zeta}_{t^{\prime},s^{\prime}}({\nabla}f^{\prime})
\bigr)=-\bigl[W({\cal L}^{\star}f,f^{\prime})+W(f,{\cal L}^{\star}f^{\prime})\bigr]
{\vert}[s,t]{\cap}[s^{\prime},t^{\prime}]{\vert}$$
$${\forall} \ f,f^{\prime}{\in}
{\cal D}^{n}({\Omega}), \  t,s,t^{\prime},s^{\prime}{\in}{\bf R},\eqno(3.7)$$
where the last factor is the length of the intersection of the intervals $(s,t)$ and 
$(s^{\prime},t^{\prime})$.
\vskip 0.3cm
\centerline {\bf 3.2. The Chaoticity Hypothesis.}
\vskip 0.3cm
We assume that the stochastic current is chaotic in the  sense that the space-time 
correlations of the unsmeared stochastic field ${\zeta}_{t,s}^{stoc}$ are of short range 
on the microscopic scale and therefore, since $L_{N}$ tends to infinity with $N$, of zero 
range on the hydrodynamic scale. Further, in accordance with the central limit theorem 
for fluctuation fields with short range correlations [21], we assume that the process 
${\zeta}^{stoc}$ is Gaussian.  Thus our chaoticity assumption is that
\vskip 0.2cm\noindent
(C.1) the process ${\zeta}^{stoc}$ is Gaussian; and
 \vskip 0.2cm\noindent
(C.2)$E\bigl({\zeta}_{t,s}^{stoc}(g){\zeta}_{t^{\prime},s^{\prime}}(g^{\prime})\bigr)
=0$ if either $(s,t){\cap}(s^{\prime},t^{\prime})$ or ${\rm supp}(g){\cap}
{\rm supp}(g^{\prime})$ is empty.
\vskip 0.2cm
The following proposition was inferred\footnote*{In fact the proof in [4] invoked the 
supplementary assumption that the l.h.s of Eq. (3.8) was continuous in its time variables. 
However, that assumption is redundant, since the required continuity can be derived by 
exploiting the positivity of that formula for $f=f^{\prime}, \ t=t^{\prime}, \ 
s=s^{\prime}$.} in [4] from this assumption and Schwartz\rq s compact and point 
support theorems [20, Ths. 26,35]. 
\vskip 0.3cm
{\bf Proposition 3.1.} {\it Under the assumption (C.2), the two-point function for 
${\zeta}$ takes the form
$$E\bigl({\zeta}_{t,s}(g){\zeta}_{t^{\prime},s^{\prime}}(g^{\prime})\bigr)=
Z(g,g^{\prime}){\vert}[s,t]{\cap}[s^{\prime},t^{\prime}]{\vert} \ {\forall} \ 
g,g^{\prime}{\in}{\cal D}^{n}({\Omega}), \ t,s,t^{\prime},s^{\prime}{\in}{\bf R},
\eqno(3.8)$$
where $Z$ is an element of ${\cal D}^{{\prime}n}({\Omega}){\otimes}
{\cal D}^{{\prime}n}({\Omega})$ with support in the domain 
${\lbrace}(x,x){\vert}x{\in}{\Omega}{\rbrace}$.} 
\vskip 0.3cm
The following corollary to this proposition follows immediately  from a comparison of 
Eqs. (3.7) and (3.8).
\vskip 0.3cm
{\bf Corollary 3.2.} {\it Under the same assumptions, 
$$Z({\nabla}f,{\nabla}f^{\prime})=-\bigl[W({\cal L}^{\star}f,f^{\prime})+
W(f,{\cal L}f^{\prime})\bigr] \ {\forall} \ f,f^{\prime}{\in}{\cal D}^{n}({\Omega}).
\eqno(3.9)$$}
\vskip 0.3cm
\centerline {\bf  3.3. The Local Equilibrium Hypothesis.} 
\vskip 0.3cm
For an {\it equilibrium} state, the static two point function for the field ${\xi}$ has been 
derived from the KMS condition in the following form  [2, Ch. 7, Appendix C].
$$W_{eq}(f,f^{\prime}){\equiv}E_{eq}\bigl({\xi}_{eq}(f){\xi}_{eq}(f^{\prime})\bigr)
=\bigl(f,J_{q}f^{\prime}\bigr) \ {\forall} \ 
f,f^{\prime}{\in}{\cal D}^{n}({\Omega}),\eqno(3.10)$$
where the subscript $eq$, here and elsewhere, refers to the equilibrium state and the 
function $J_{q}$ and the inner product $(.,.)$ and are defined in Eqs. (2.1) and (2.7), 
respectively. Further, the process ${\xi}_{eq}$ inherits from its underlying quantum 
dynamics the property of invariance under time reversals. Hence, in view of the 
stationarity of this process, it follows from a standard argument [4] that
$$W_{eq}({\cal L}_{eq}^{\star}f,f^{\prime})=W_{eq}
({\cal L}_{eq}^{\star}f^{\prime},f) \ {\forall} \ f,f^{\prime}
{\in}{\cal D}^{n}({\Omega}).\eqno(3.11)$$
\vskip 0.2cm
Turning now to the process ${\zeta}^{stoc}$, we see that, by translational invariance of 
the equilibrium state, it follows from by Eqs. (2.4)-(2.6) that ${\cal 
L}_{eq}=K_{q}{\Delta}$ and thence, by Eqs. (2.7), (3.9) and (3.10), that
$$Z_{eq}({\nabla}f,{\nabla}f^{\prime})=
\bigl({\nabla}f,[{\tilde K}_{q}+{\tilde K}_{q}^{\star}]
{\nabla}f^{\prime}\bigr)_{V},\eqno(3.12)$$
where ${\tilde K}_{q}^{\star}$ is the transpose of the matrix ${\tilde K}_{q}$. Further, 
a simple argument based on the rotational symmetry of the equilibrium state leads to an 
extension of this formula to the following one [4]. 
$$Z_{eq}(g,g^{\prime})=
\bigl(g,[{\tilde K}_{q}+{\tilde K}_{q}^{\star}]g^{\prime}\bigr)_{V} \ {\forall} \ 
g,g^{\prime}{\in}{\cal D}_{V}^{n}({\Omega}).\eqno(3.13)$$
\vskip 0.2cm
Eqs. (3.10), (3.11) and (3.13) are our equilibrium conditions for the fluctuation process.
We now formulate the local properties of these conditions in terms of test functions that 
are highly localised around an arbitrary point $x_{0}$ of ${\Omega}$. Thus, for 
$x_{0}{\in}{\Omega}$ and sufficiently small ${\epsilon}{\in}{\bf R}_{+}$, we define 
the transformations $f{\rightarrow}f_{x_{0},{\epsilon}}$ of  
${\cal D}^{n}({\Omega})$ and $g{\rightarrow}g_{x_{0},{\epsilon}}$ of  
${\cal D}_{V}^{n}({\Omega})$ by the formulae
$$f_{x_{0},{\epsilon}}(x)={\epsilon}^{-d/2}f\bigl({\epsilon}^{-1}(x-x_{0})\bigr);
 \ g_{x_{0},{\epsilon}}(x)={\epsilon}^{-d/2}g\bigl({\epsilon}^{-1}(x-x_{0})\bigr).
\eqno(3.14)$$
Thus, since by Eqs. (2.4)-(2.6), ${\cal L}_{eq}=K_{q}{\Delta}$ and thus 
${\epsilon}^{2}{\cal L}_{eq}^{\star}f_{x_{0},{\epsilon}}=[{\cal 
L}^{\star}f]_{x_{0},{\epsilon}}$, it follows that the equilibrium conditions (3.10), 
(3.11) and (3.13) are equivalent to the following ones, which evidently represent 
properties at the (hydrodynamical) point $x_{0}$ in the limit 
${\epsilon}{\rightarrow}0$. 
$$W_{eq}(f_{x_{0},{\epsilon}},
f_{x_{0},{\epsilon}}^{\prime})=\bigl(f,J_{q}f^{\prime}\bigr) \ {\forall} \ 
f,f^{\prime}{\in}{\cal D}^{n}({\Omega}), \ x_{0}{\in}{\Omega},\eqno(3.15)$$
$$ {\epsilon}^{2}W_{eq}
({\cal L}_{eq}^{\star}f_{x_{0},{\epsilon}},f_{x_{0},{\epsilon}}^{\prime})=
{\epsilon}^{2}W_{eq}
({\cal L}_{eq}^{\star}f_{x_{0},{\epsilon}}^{\prime},f_{x_{0},{\epsilon}}) 
\ {\forall} \ f,f^{\prime}{\in}{\cal D}^{n}({\Omega}), \ 
x_{0}{\in}{\Omega},\eqno(3.16)$$
both sides of this equation being equal to $({\cal L}_{eq}^{\star}f,J_{q}f^{\prime})$; 
and
$$Z_{eq}(g_{x_{0},{\epsilon}},
g_{x_{0},{\epsilon}}^{\prime})=
\bigl(g,[{\tilde K}_{q}+{\tilde K}_{q}^{\star}]g^{\prime}\bigr)_{V} \ {\forall} \ 
g,g^{\prime}{\in}{\cal D}_{V}^{n}({\Omega}), \ x_{0}{\in}{\Omega}\eqno(3.17)$$
Correspondingly, we take our local equilibrium conditions for the nonequilibrium steady 
state to be given by removing the subscript $eq$, replacing $J_{q}$ and $K_{q}$ in 
these formulae by $J_{q}(x_{0})$ and $K_{q}(x_{0})$, respectively, and then passing 
to the limit ${\epsilon}{\rightarrow}0$. Thus, the resultant conditions are
$${\rm lim}_{{\epsilon}{\downarrow}0}W(f_{x_{0},{\epsilon}},
f_{x_{0},{\epsilon}}^{\prime})=\bigl(f,J_{q}(x_{0})f^{\prime}\bigr) \ {\forall} \ 
f,f^{\prime}{\in}{\cal D}^{n}({\Omega}), \ x_{0}{\in}{\Omega},\eqno(3.18)$$
$${\rm lim}_{{\epsilon}{\downarrow}0}{\epsilon}^{2}W
({\cal L}^{\star}f_{x_{0},{\epsilon}},f_{x_{0},{\epsilon}}^{\prime})=
{\rm lim}_{{\epsilon}{\downarrow}0}{\epsilon}^{2}W
({\cal L}^{\star}f_{x_{0},{\epsilon}}^{\prime},f_{x_{0},{\epsilon}}) 
\ {\forall} \ f,f^{\prime}{\in}{\cal D}^{n}({\Omega}), \ 
x_{0}{\in}{\Omega},\eqno(3.19)$$
and
$${\rm lim}_{{\epsilon}{\downarrow}0}Z(g_{x_{0},{\epsilon}},
g_{x_{0},{\epsilon}}^{\prime})=\bigl(g,[{\tilde K}_{q}(x_{0})+{\tilde 
K}_{q}^{\star}(x_{0})] g^{\prime}\bigr)_{V} \ {\forall} \ g,g^{\prime}{\in}{\cal 
D}_{V}^{n}({\Omega}), \ x_{0}{\in}{\Omega}.\eqno(3.20)$$
Further, by Eq. (3.8),  the chaoticity assumption (C.2) implies that $Z(g,g^{\prime})$ 
vanishes if the supports of $g$ and $g^{\prime}$ are disjoint. Consequently, by
Schwartz\rq s compact and point support theorems [20, Ths. 26,35], Eq. (3.20) implies 
that [4]
$$Z(g,g^{\prime})=(g,[{\tilde K}_{q}+{\tilde K}_{q}^{\star}]g^{\prime})_{V} \ 
{\forall} g,g^{\prime}{\in}{\cal D}_{V}({\Omega}).\eqno(3.21)$$
Hence, by Eq. (3.7),
$$E\bigl({\zeta}_{t,s}^{stoc}(g){\zeta}_{t^{\prime},s^{\prime}}(g^{\prime})\bigr)=
(g,[{\tilde K}_{q}+{\tilde K}_{q}^{\star}]g^{\prime})_{V}
{\vert}[s,t]{\cap}[s^{\prime},t^{\prime}]{\vert} \  
{\forall} \ g,g^{\prime}{\in}{\cal D}_{V}^{n}({\Omega}), \ t,s,t^{\prime},s^{\prime}
{\in}{\bf R}.\eqno(3.22)$$
\vskip 0.3cm
\centerline {\bf 3.4. Generalised Reciprocity Relations and Onsager-Machlup Processes.}
\vskip 0.3cm
The following proposition was proved in [4].
\vskip 0.3cm
{\bf Proposition 3.3.} {\it Assuming the regression hypotheses and local equilibrium 
conditions (3.18) and (3.19), the $n$-by-$n$ matrix-valued function ${\tilde K}_{q}$ on 
${\Omega}$ is symmetric. This constitutes a nonequilibrium, position-dependent 
generalisation of the Onsager reciprocity relations.}
\vskip 0.3cm
In order to show that the Onsager-Machlup theory extends to the present setting, we 
define 
$$w_{t}(f)={\zeta}_{t,0}^{stoc}({\nabla}f)  \ {\forall} \ f{\in}{\cal D}^{n}({\Omega}), 
\ t{\in}{\bf R}\eqno(3.23)$$
and infer from  this formula, together with Eq. (3.22), Prop. 3.1 and the Gaussian 
condition (C.1) that $w$ is the Wiener process whose two-point function is 
$$E\bigl([w_{t}(f)-w_{s}(f)][w_{t^{\prime}}(f^{\prime})-
w_{s^{\prime}}(f^{\prime})]\bigr)=2({\nabla}f,{\tilde K}_{q}{\nabla}f^{\prime})_{V} 
\ {\forall} \ f,f^{\prime}{\in}{\cal D}^{n}({\Omega})_{V}, \ 
t,s,t^{\prime},s^{\prime}{\in}{\bf R}.\eqno(3.24)$$
The following proposition then ensues immediately from Eqs. (3.6), (3.23) and (3.24). 
\vskip 0.3cm
{\bf Proposition 3.4.} {\it Assuming the regression, chaoticity and local equilibrium 
hypothese, the ${\xi}$-field executes a generalised Onsager-Machlup process given by 
the Langevin equation
$$d{\xi}_{t}={\cal L}{\xi}_{t}dt+dw_{t},\eqno(3.25)$$
where $w$ is the Wiener process specified above.}
\vskip 0.5cm
\centerline {\bf  4. Long Range Correlations of the ${\xi}$-Process.}
\vskip 0.3cm
The next proposition and the subsequent comment signify that, under the above 
assumptions, the spatial correlations of the ${\xi}$ field are generically of non-zero range 
on the macroscopic scale and hence of long (infinite!) range on the microscopic one.
\vskip 0.3cm
{\bf Proposition 4.1.} {\it Let ${\Phi}_{q}$ and ${\Psi}_{q}$ be the be the $n$-by-$n$ 
matrix valued scalar and vector fields, respectively, in ${\Omega}$ defined by the 
formula
$${\Phi}_{q}={\Delta}{\tilde K}_{q}+{\nabla}.{\Psi}_{q} \eqno(4.1)$$
and
$${\Psi}_{q;k,l}(x)={\sum}_{k^{\prime},l^{\prime}=1}^{n}
\bigl[{{\partial}\over {\partial}q_{l^{\prime}}(x)}
K_{k,k^{\prime}}\bigl(q(x)\bigr)\bigr]
\bigl[J_{l,l^{\prime}}\bigl(q(x)\bigr){\nabla}q_{k^{\prime}}(x)-J_{k^{\prime},l}
\bigl(q(x)\bigr){\nabla}q_{l^{\prime}}(x)\bigr].\eqno(4.2)$$
Then, under the assumptions of the regression, chaoticity and local equilibrium 
hypotheses, a sufficient condition for the static spatial correlations of the ${\xi}$-process 
to be of non-zero range on the macroscopic scale is that either ${\Phi}_{q}$ does not 
vanish or that the matrix ${\Psi}_{q}$ is symmetric.} 
\vskip 0.3cm
{\bf Comment.} The functions ${\Phi}_{q}$ and ${\Psi}_{q}$ are determined by the 
forms of the entropy function $s$, the transport coefficient $K$ and the form of 
$q_{{\vert}{\partial}{\Omega}}$, which represents the boundary conditions imposed by 
the reservoirs.on the profile of the function $q$. Since these functions are mutually 
independent, it follows from Prop. 4.1 that it is only under very special conditions that the 
spatial correlations of the ${\xi}$-field are of zero range on the hydrodynamic scale. In 
other words, these correlations are generically of non-zero range on that scale and 
consequently of infinite range on the microscopic one. 
\vskip 0.3cm
{\bf Example.} In the case of the simple exclusion process, whether in the classical 
picture [8-10] or the quantum one [7], $n=1, \ d=1, \ K(q)=1, \ s(q)=-q{\rm log}(q)-(1-
q){\rm log}(1-q)$ and $q(x)=a+bx$, where $a$ and $b$ are constants, the latter being 
non-zero in nonequilibrium steady states. Then it follows from Eqs. (2.1), (4.1) and (4.2) 
that, in these states, ${\Psi}_{q}=0$ and ${\Phi}_{q}(x)=-2b^{2}{\neq}0$. Hence, long 
(infinite) range correlations prevail in this model.
\vskip 0.3cm
{\bf Proof of Prop. 4.1}. We employ a {\it reductio ad absurdum} argument. Thus we 
start by assuming that the spatial correlations of the ${\xi}$-process are of zero range, on 
the hydrodynamic scale, i.e. that the support of $W$ lies in the domain 
${\lbrace}x,x^{\prime}{\in}{\Omega}{\vert}x=x^{\prime}{\rbrace}$. Then by pursuing 
the argument used to obtain Eq. (3.21) , we see that this assumption, together with the 
local equilibrium condition (3.18), implies that
$$W(x,x^{\prime})=J_{q}(x){\delta}(x-x^{\prime}).\eqno(4.3)$$
On the other hand, Eq. (3.21) and Prop. 3.3 imply that Eq. (3.9) is equivalent to the 
following equation for the distribution $W$.
$$[{\cal L}+{\cal L}^{\prime}]W(x,x^{\prime})=
2{\nabla}.\bigl({\tilde K}_{q}(x){\nabla}{\delta}(x-x^{\prime})\bigr),\eqno(4.4)$$
where ${\cal L}^{\prime}$ is the version of ${\cal L}$ that acts on functions of 
$x^{\prime}$. On inserting this formula (4.3) for $W$ into this equation and using Eqs. 
(2.4)-(2.6), , it follows, after some manipulation, that
$${\Phi}_{q}(x){\delta}(x-x^{\prime})+\bigl[{\Psi}_{q}(x)-
{\Psi}_{q}^{\star}(x)\bigr]{\delta}^{\prime}(x-x^{\prime})=0.$$
Thus, the assumption that the spatial correlations of the ${\xi}$-process are of zero range 
on the hydrodynamic scale cannot be sustained unless ${\Phi}_{q}$ vanishes and 
${\Psi}_{q}$ is symmetric.
\vskip 0.5cm
\centerline {\bf 5. Concluding Remarks.} 
\vskip 0.3cm
In this work, the crucial links between quantum microdynamics and classical 
macroscopic continuum mechanics  are provided by the mesoscopic picture of 
hydrodynamical fluctuations. Apart from generalising both the Onsager reciprocity 
relations and the Onsager-Machlup fluctuation process to nonequilibrium steady states, 
the ensuing theory establishes, on a very general basis, that the spatial correlations of the 
hydrodynamical variables are generically of infinite range in these states. This result, 
which had previously been established for certain classical stochastic models [8-10], 
marks a crucial difference between equilibrium and nonequilibrium steady states, since 
the spatial correlations of the former are of short range, except at critical points [22].
\vskip 0.5cm
\centerline {\bf References.}
\vskip 0.3cm\noindent
[1] L.Onsager: Phys. Ref. {\bf 37}, 405, 1931; Phys. Rev. {\bf 38}, 2265, 1931.
\vskip 0.2cm\noindent
[2] G. L. Sewell: {\it Quantum Mechanics and its Emergent Macrophysics}, Princeton 
Univ. Press, Primceton, 2002.
\vskip 0.2cm\noindent
[3] G. L. Sewell: Lett. Math. Phys. {\bf 68}, 53, 2004.
\vskip 0.2cm\noindent
[4] G. L. Sewell: Rev. Math. Phys. {\bf 17}, 977, 2005.
\vskip 0.2cm\noindent
[5] G. L. Sewell: Helv. Phys. Acta {\bf 67}, 4, 1994.
\vskip 0.2cm\noindent
[6] B. Nachtergaele and H.-T. Yau: Commun. Math. Phys. {\bf 243}, 485, 2003.
\vskip 0.2cm\noindent
[7] G. L. Sewell: Rep. Math. Phys. {\bf 59}, 243, 2007. 
\vskip 0.2cm\noindent
[8] H. Spohn: J. Phys. A {\bf 16}, 4275, 1983
\vskip 0.2cm\noindent
[9] B. Derrida, J. L. Lebowitz and E. R. Speer: J. Stat. Phys. {\bf 107}, 599, 2002.
\vskip 0.2cm\noindent
[10] L. Bertini, A. de Sole, D. Gabrielli, G. Jona-Lasinio and C. Landim: J. Stat. Phys. 
{\bf 107}, 635, 2002.
\vskip 0.2cm\noindent
[11] L. Onsager and S. Machlup: Phys. Rev. {\bf 91}, 1505, 1953.
\vskip 0.2cm\noindent
[12] D. Ruelle: {\it Statistical Mechanics}, Benjamin, New York, 1969.  
\vskip 0.2cm\noindent
[13] G. G. Emch: {\it Algebraic Methods in Statistical mechanics and Quantum Field 
Theory}, Wiley, New York, 1971.
\vskip 0.2cm\noindent
[14] W. Thirring: {\it Quantum Mechanics of Large Systems}, Springer, New York, 
1983.
\vskip 0.2cm\noindent
[15] R. Haag, N. M. Hugenholtz and M. Winnink: Commun. Math. Phys. {\bf 5}, 215, 
1967.
\vskip 0.2cm\noindent
[16] G. G. Emch, H. J. F. Knops and E. Verboven: J. Math. Phys. {\bf 11}, 1655, 1970.
\vskip 0.2cm\noindent
[17] D. Ruelle: J. Stat. Phys. {\bf 98}, 57, 2000.
\vskip 0.2cm\noindent
[18]  S. Tasaki and T. Matsui: Pp. 100-119 of {\it Fundamental Aspects of Quantum 
Theory}, Ed. L.Accardi and S. Tasaki, World Scientific, Singapore, 2000.
\vskip 0.2cm\noindent
[19] R. F. Streater and A. S. Wightman: {\it PCT, Spin and Statistics and All That}, 
Benjamin, New York, 1964.
\vskip 0.2cm\noindent
[20] L.Schwartz: {\it Theorie des Distributions}, Hermann, Paris, 1998.
\vskip 0.2cm\noindent
[21] D. Goderis, P. Vets and A. Verbeure: Prob. Theory Related Fields {\bf 82}, 527, 
1989. 
\vskip 0.2cm\noindent
[22] M. Broidio, B. Momont and A. Verbeure: J. Math. Phys. {\bf 36}, 6746, 1995.
\end